\documentclass{ifacconf}
\usepackage{url}
\usepackage{graphicx}      
\usepackage{natbib}   
\usepackage{amsmath}
\usepackage{mathtools}
\usepackage{amsfonts}
\usepackage{lipsum}
\usepackage{xcolor}
\usepackage{tikz}
\usepackage[percent]{overpic} 

\usepackage{fancyhdr}

\fancypagestyle{firstpage}
{   
    \fancyhead[C]{This work has been submitted to IFAC for possible publication}
    \fancyhead[R]{}
    \fancyfoot[C]{}
}

\DeclareMathOperator*{\argmin}{arg\,min}

\definecolor{draftcolor}{RGB}{125,125,200}
\definecolor{k}{HTML} {000000}
\definecolor{bluePoli}{HTML} {003576}
\definecolor{orPoli}{HTML} {e57200}
\definecolor{greenPoli}{HTML} {008148}
\definecolor{grey}{HTML} {aaaaaa}
\definecolor{purMAT}{rgb} {0.4940 0.1840 0.5560}
\definecolor{green2}{HTML}{2c9c3a}

\newtheorem{remark}{\noindent \textbf{Remark}}{\normalfont}{\normalfont}

\newlength{\myLineWidth}
\newlength{\myLineLength}
\newsavebox{\myGreenBox}
\savebox{\myGreenBox}{\tikz{\draw[greenPoli,solid,line width=\myLineWidth](0,0) -- (\myLineLength,0);}}

\newsavebox{\myOrBox}
\savebox{\myOrBox}{\tikz{\draw[orPoli,solid,line width=\myLineWidth](0,0) -- (\myLineLength,0);}}
\newcommand{\lOr}{\raisebox{2pt}{\usebox{\myOrBox}}}

\newsavebox{\myBlueBox}
\savebox{\myBlueBox}{\tikz{\draw[bluePoli,solid,line width=\myLineWidth](0,0) -- (\myLineLength,0);}}
\newcommand{\lBlue}{\raisebox{2pt}{\usebox{\myBlueBox}}}

\newsavebox{\myGreyBox}
\savebox{\myGreyBox}{\tikz{\draw[grey,solid,line width=\myLineWidth](0,0) -- (\myLineLength,0);}}
\newcommand{\lGrey}{\raisebox{2pt}{\usebox{\myGreyBox}}}

\newsavebox{\myKBox}
\savebox{\myKBox}{\tikz{\draw[k,solid,line width=\myLineWidth](0,0) -- (\myLineLength,0);}}
\newcommand{\lK}{\raisebox{2pt}{\usebox{\myKBox}}}

\let\oldlipsum\lipsum
\renewcommand{\lipsum}[1][]{%
  {\color{draftcolor}\oldlipsum[#1]}%
}

\usepackage{enumitem}
 
\begin{document}
\begin{frontmatter}

\title{Unifying Decision Making and Trajectory Planning in Automated Driving through Time-Varying Potential Fields\thanksref{footnoteinfo}}

\thanks[footnoteinfo]{This publication is part of the project Piano Nazionale di Ripresa e Resilienza (PNRR)- Next Generation Europe, which has received funding from the Italian Ministry of University and Research – DM 117/2023}

\author[First]{D. Costa}
\author[First]{F. Cerrito} 
\author[First]{M. Canale} 
\author[First]{C. Novara}
\vspace{-0.2cm}
\address[First]{Politecnico di Torino, Torino 10129, Italy.}
\vspace{-0.2cm}
\begin{abstract}                
This paper proposes a unified decision making and local trajectory planning framework based on Time-Varying Artificial Potential Fields (TVAPFs). The TVAPF explicitly models the predicted motion via bounded uncertainty of dynamic obstacles over the planning horizon, using information from perception and V2X sources when available. TVAPFs are embedded into a finite horizon optimal control problem that jointly selects the driving maneuver and computes a feasible, collision free trajectory. The effectiveness and real-time suitability of the approach are demonstrated through a simulation test in a multi-actor scenario with real road topology, highlighting the advantages of the unified TVAPF-based formulation.
\end{abstract}
\begin{keyword}
Trajectory and path planning for AVs, Autonomous vehicles, Control architectures in automotive control.
\end{keyword}

\end{frontmatter}

\section{Introduction}\vspace{-0.3cm}
\thispagestyle{firstpage} 
The continuous introduction of Advanced Driver Assistance Systems~(ADAS), such as Anti-lock Braking Systems~(ABS), Electronic Stability Programs~(ESP), and automatic emergency braking, has significantly contributed to improving vehicle safety and reducing crash severity. Building upon this foundation, the development of Autonomous Vehicles~(AVs) promises to further enhance the safety, efficiency, and accessibility of transportation. 
A technical challenge in this domain is the creation of reliable control systems capable of on-board, real-time Trajectory Planning (TP). This planning must ensure that the vehicles reach the destination while strictly adhering to traffic laws, maintaining passenger comfort, respecting vehicle dynamics limitations, and handling interactions with uncertain external actors.\\
A significant challenge in TP for AVs lies in the common separation between high-level decision making (DM) (e.g., performing an overtaking maneuver, maintaining the current lane, initiating braking) and low-level trajectory planning (generating a smooth path and velocity profile). This hierarchical approach is widely adopted, with works such as \cite{FSM1} proposing a supervisory Finite State Machine ($\text{FSM}$) for DM in emergency conditions, and \cite{SWctrl} introducing a switching control approach for two-lane country roads (c.f. \cite{survDM}). Despite these methods performing well in simplified or structured situations, they face considerable limitations in complex multi-actor scenarios. This is primarily because hierarchical systems tend to address one primary task at a time. In real-world driving, however, the conditions are often a superimposition of multiple events (e.g., overtake in the presence of a slow leading vehicle and/or an oncoming vehicle in the opposite lane). A rigid FSM design can struggle to properly resolve possible conflicting objectives. This can lead to behavioral logic failures due to unforeseen state combinations that were not explicitly accounted for during the design phase. \\
To address this complexity, other approaches, e.g., \cite{ColAvoid} and \cite{Bezier}, decouple the problem by defining a suitable path and a proper speed profile independently. While this technique simplifies the overall problem resolution and may guarantee the existence of a suitable path, it remains prone to critical limitations. For instance, an independent high-level decision to maintain a high speed might generate a low-level path that is immediately infeasible due to physical vehicle constraints or high collision risk. Consequently, unifying the DM and TP steps within a single integrated framework is highly desirable (c.f. \cite{DMLTP, DMLTP2}). This unification guarantees that the planned trajectory is inherently safe, feasible, and consistent with the intended driving strategy, despite the uncertain and evolving behavior of surrounding actors.\\
A widely adopted approach for real-time obstacle avoidance and path planning is the Artificial Potential Field (APF) method, originally introduced by \cite{APF0}. Inspired by physical potential fields, this framework typically superimposes repulsive fields onto obstacles or hazardous areas, while attractive fields guide the vehicle toward the target configuration. Standard APF formulations characterize external obstacles using \emph{static} shapes that do not evolve over time, accounting only for the instantaneous geometry of the environment while neglecting dynamic evolution and state uncertainty. In the context of path planning, APFs are primarily utilized within gradient-based schemes to generate virtual forces that steer the vehicle toward the goal, see, e.g., \cite{APF_std0}. Specific field shapes are employed to ensure collision avoidance, and various techniques have been developed to prevent entrapment in local minima and enhance overall performance \citep{APF_std1}.\\
However, significant limitations arise when addressing the dynamic and stochastic nature of external actors in real-world scenarios. Static APFs inherently struggle to account for uncertainties regarding the future state evolution (positions and velocities) of other traffic participants.  Such uncertainties stem from imprecise or absent Vehicle-to-Everything (V2X) communication regarding future intent and infrastructure delayed updates. Furthermore, the complete absence of V2X communication necessitates that the system explicitly consider the full range of reachable states for surrounding actors.\\
To mitigate these issues, several \emph{adaptive} approaches have been proposed. For instance, \cite{AAPF} introduced Adaptive Potential Fields that incorporate V2X data, such as heading and steering angles, to estimate the intentions of surrounding vehicles. Similarly, \cite{AAPF1} exploited acceleration and mass information to refine path planning. While these adaptations improve upon standard static APFs, they generally lack inherent guarantees of trajectory safety with respect to the future behavior of surrounding actors.\\
To address these challenges, this work proposes a \emph{Time-Varying} Artificial Potential Field (TVAPF) formulation. The TVAPF provides a unified framework capable of directly integrating and accounting for the aforementioned uncertainties, regardless of their origin (e.g., V2X data or sensor fusion outputs). In this regard, the APF shape is not static; instead, it evolves over the Local Trajectory Planner (LTP) prediction horizon, exploiting all available information to dynamically model the environment. By incorporating the TVAPF within a suitable Finite Horizon Optimal Control Problem (FHOCP), the system simultaneously determines the optimal driving maneuver and the corresponding physically feasible trajectory. This approach effectively unifies DM and TP problems, yielding real-time and strategically coherent control actions.
The main contributions of this paper are summarized as follows:
\begin{enumerate}[leftmargin=*]
    \item The definition and development of a novel TVAPF formulation that explicitly accounts for the state evolution and associated uncertainties of dynamic obstacles.
    \item The unification of the DM and the TP problem through a tailored FHOCP, ensuring the generation of reliable trajectories.
    \item The validation and the computational effectiveness of the proposed method in a challenging driving scenario, with real road topologies and multi-actor dynamics, confirms the efficacy of the unified framework for real-time applications.
\end{enumerate}
\subsubsection{Notation}
    The set of integers over a specific interval $[0, a]$ is denoted as: $\mathbb{N}_{a} = \{i \in \mathbb{N} \mid 0 \leq i \leq a\}$.
    FHOCP objective functions are denoted by a boldface letter $\mathbf{F_{\mathrm{obj}}}$. The standard FHOCP notation is used to denote predictions. The predicted value $x$ at time instant $k+j$, computed using information available at the current time instant $k$, is denoted by $x(k+j|k)$. Given the condition $P=P^\top\succeq 0$, which denotes that $P$ is a symmetric Positive Semi-Definite matrix, the weighted $P$-norm of a vector $x$ is defined as $\lVert x \rVert_{P} = \sqrt{x^\top P x}$.
\section{Problem set-up}\label{sec:1}\vspace{-0.3cm}
This section introduces the coordinate reference frames used in the development of the proposed approach. Subsequently, the comprehensive AV control architecture is presented, detailing its subsystems and their functional interconnections.
\vspace{-0.1cm}\subsection{Reference Frames}\label{sec:fre} 
\begin{figure}[]
    \centering
    \includegraphics[width=0.6\linewidth]{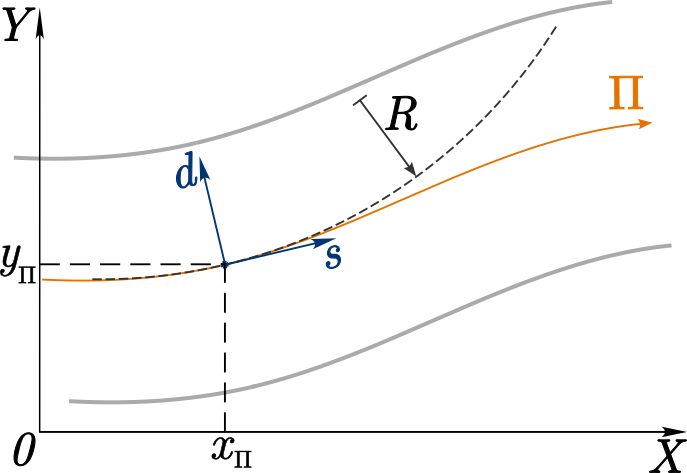}
    \caption{Architecture reference frames. \lK Cartesian coordinates, \lBlue Frenet-Serret coordinates, \lOr GPP reference path $\Pi$, and \lGrey Road boundaries.}
    \label{fig:freRF}
\end{figure}

\vspace{-0.3cm}
To facilitate the development of the control architecture, two distinct reference frames are employed, as illustrated in Fig.~\ref{fig:freRF}.\\
The first is a global Cartesian frame defined by the $(X, Y)$ axes.  This frame represents the natural choice for describing the external environment, allowing for direct definitions of Euclidean distances and absolute road geometries.\\
Conversely, to effectively define and manage geometric constraints related to road boundaries, curvature, and lane positioning, the LTP problem is formulated using a local Frenet reference frame. In this frame, the state of a generic AV is described by the longitudinal coordinate $s$ (representing the curvilinear distance along the reference path) and the lateral offset $d$ (representing the perpendicular distance from the path). This coordinate system, often referred to as Frenet-Serret coordinates, effectively decouples the longitudinal motion along the path from lateral deviations, making it the preferred choice for describing vehicle motion in structured road environments.
\begin{remark}\label{rm:FC}
    Given a regular path curve belonging to the class $\mathcal{C}^2$, a bijective analytical transformation $\mathcal{T}_{\text{cf}}$ exists to map coordinates from the Cartesian frame to the Frenet frame, and its inverse $\mathcal{T}_{\text{fc}}$ is consistently defined (cf \cite{Bezier}).
\end{remark}\vspace{-.2cm}
\vspace{-0.1cm}\subsection{Control architecture}
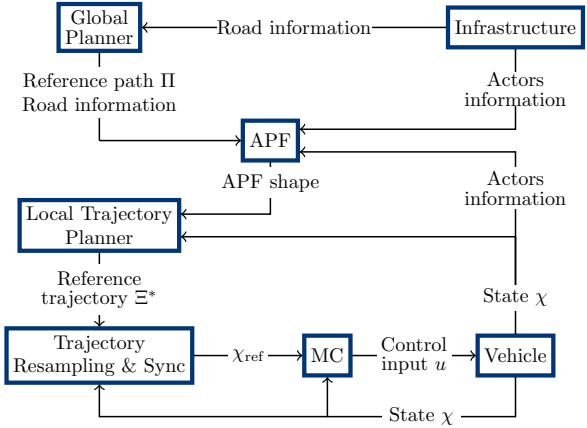
\begin{figure}
    \centering
    \begin{tikzpicture}[every path/.style={line width=.2mm},scale=0.75, transform shape]
\tikzstyle{block} = [draw, rectangle, minimum height=2em, minimum width=2em, ultra thick]
    \node (GP) [block, draw=bluePoli] {\shortstack{Global\\ Planner}};
    \node (TVAPF) [block, below of=GP, yshift=-1cm, xshift=3cm, draw=bluePoli] {APF};
    \node (LPP) [block, below of=GP, yshift=-2.5cm, draw=bluePoli] {\shortstack{Local Trajectory \\ Planner}};
    \node (RefT) [block, below of=LPP, yshift=-1.3cm, draw=bluePoli] {\shortstack{Trajectory \\ Resampling \& Sync}};
    \node (NMPC) [block, right of=RefT, xshift=3cm, draw=bluePoli] {MC};
    \node (VH) [block, right of=NMPC, xshift=2.3cm, draw=bluePoli] {Vehicle};
    \node (IF) [block, right of=GP, xshift=6.3cm, draw=bluePoli] {\shortstack{Infrastructure}};
    \coordinate (mid1) at ([yshift=-2cm] GP);
    \draw [-] (GP) -- (mid1) node[pos=0.45, fill=white] {\shortstack[]{Reference path $\Pi$ \\ Road information}};
    \draw[->] (TVAPF) |- ([yshift=0.2cm]LPP.east)node[pos=0.2, fill=white]{APF shape};
    \draw[->] (LPP) -- (RefT)node[midway, fill=white]{\shortstack{Reference \\ trajectory $\Xi^*$}};
        \draw[->] (RefT)--(NMPC)node[midway, fill=white, inner sep=2pt]{$\chi_{\text{ref}}$};
    \draw[->] (NMPC)--(VH)node[midway, fill=white, inner sep=2pt]{\shortstack{Control \\ input $u$}};
    \draw[->] (IF)--(GP)node[midway, fill=white, inner sep=0pt]{Road information};
    \draw[->] (IF)|-([yshift=0.2cm]TVAPF.east)node[near start, fill=white, inner sep=0pt]{\shortstack[]{Actors \\information}};
    \draw[->] (VH)|-([yshift=-0.2cm]LPP.east)node[pos=0.1, fill=white, inner sep=9pt]{};
    \draw[->] (VH)|-([yshift=-0.2cm]TVAPF.east)node[pos=0.1, fill=white, inner sep=2pt]{\shortstack{State $\chi$}}node[pos=0.4, fill=white, inner sep=2pt]{\shortstack{Actors \\information}};
    \draw[->](mid1)--(TVAPF);

    \coordinate (x) at ([yshift=-0.7cm]VH.south);  
    \coordinate (y) at ([yshift=-0.7cm]NMPC.south);             
    \draw [->](VH.south) -- (x) -| (NMPC.south)node[pos=0.25, fill=white]{\shortstack{State $\chi$}};  
    \draw[->](y) -|(RefT.south);
\end{tikzpicture}
    \caption{Control structure architecture and exchanged information.}
    \label{fig:Arch}
\end{figure}

\vspace{-0.3cm}
This paper proposes a flexible LTP for AVs. 
While the focus of this work is on the LTP, its design is critically dependent on a systematic and complete view of the full control architecture. This structure is pivotal for properly identifying how information is collected and processed across different functionalities, thereby ensuring a well-posed control problem.
The proposed control architecture (Fig.~\ref{fig:Arch}) is presented in this section, providing details on both the implementation of each layer and how information is collected and shared with lower-level control blocks.
\begin{itemize}[leftmargin=*]
    \item The \emph{Global Path Planner} (GPP) computes the route to the destination based on a priori and infrastructure information, generating a reference geometric path characterized by edges and lane curvature information. Additional road attributes, such as the number of lanes and speed limitations, are also provided to generate safe trajectories.
    The path is derived via standard methods, e.g., vision-based detection for highways, or graph-search algorithms like $A^*$ or Dijkstra (cf. \citep{ChLiYa14, 4082128}) combined with geometric smoothing (Dubins, clothoids, cf. \citep{CaCeBo24}) for unstructured areas. As these are established methods, a detailed description of the GPP is omitted in this paper. However, the geometric properties of the output are critical:
    \begin{defn}[Reference Path $\Pi$]\label{def:ref_path}
        The reference path $\Pi$ consists of a sequence of Cartesian points $[x_\Pi, y_\Pi]^\top$ (see Fig.~\ref{fig:freRF}). By construction, $\Pi \in \mathcal{C}^2$. This satisfies Remark~\ref{rm:FC}, guaranteeing the existence of the bijective transformation $\mathcal{T}_\text{cf}$ and its inverse along the path.
    \end{defn}
    \item The \emph{APF} block operates as an intermediate layer between the external environment (including GPP information, V2X communication, and vehicle sensor data) and the LTP. The main goal of this block is to obtain a unified description of the external environment capable of handling a complex multi-objective problem: maximizing passenger comfort, ensuring obstacle avoidance, guaranteeing compliance with traffic rules, and respecting vehicle dynamics. To this purpose, TVAPF enables the external actors' uncertain motion to be properly accounted for within the subsequent LTP block.   
    \item The \emph{LTP} processes inputs, including the reference path characteristics from the GPP, APFs information, and the current ego-vehicle state $\chi$. This planning task is formulated as a function of the APFs information within a FHOCP.    
    The solution to this FHOCP, solved with a suitable instance period, yields the optimal state trajectory $\Xi^*$ over the prediction horizon $T_L$, which the ego-vehicle tracks to execute safe, collision-free maneuvers while satisfying performance objectives.
    \item The \emph{Trajectory Resampling and Synchronization} block receives as input the optimal trajectory $\Xi^*$ computed by the LTP.  By exploiting the coordinate transformation $\mathcal{T}_\text{fc}$ and employing a linear resampling technique to account for potential mismatches between the $\Xi^*$ discretization and the motion controller sampling time, it generates a suitable reference state trajectory $\chi_{\text{ref}}$ required for the tracking control layer.
    \item A \emph{Motion Controller} (MC) computes the final low-level control actions, such as steering and acceleration, required to track the resampled optimal trajectory $\chi_{\text{ref}}$ computed by the LTP. Many well-known solutions \citep{survMC} can be exploited for this purpose; as an exemplification, a Non-linear Model Predictive Control (NMPC) is proposed in this work.
\end{itemize}
It is worth noting that, in the proposed architecture, the necessity of making a priori discrete decisions based solely on the current state of the system is eliminated. The core of the approach, the APF block, operates as a container for a unified mathematical description of both the actual state and the predicted uncertain states of the environment. Moreover, many scenarios (e.g., V2V, V2I, or absence of communication) can be handled within the same general scheme, and the architecture readily adapts to a wide range of driving situations, from high-speed highways to two-way rural roads.

\section{Trajectory Planner Objectives}~\label{sec:apf}
In this section, we first introduce the speed tracking objective and static APFs, detailing their usage. Subsequently, the focus shifts to describing the proposed TVAPF. We explain how the TVAPFs are defined, ensure safety guarantees, and enhance trajectory planning performance by exploiting all available information.
\vspace{-0.1cm}\subsection {Speed Tracking Objective} \vspace{-0.3cm}
One of the primary objectives of the LTP is the effective tracking of a desired longitudinal velocity profile. To this aim, the planner targets a desired vehicle speed $v_{\text{des}}$, which is naturally upper-bounded by the maximum allowable speed on the road segment $v_{\max}$.\\
However, the pursuit of this reference speed must account for passenger comfort, particularly concerning lateral acceleration limits. For passenger comfort, the vehicle lateral acceleration must be bounded by a threshold $a_{l,\max}$ during all maneuvers. While this constraint can be statically mapped to the road curvature, dynamic maneuvers such as lane changes require a consideration of the instantaneous angular rate of the trajectory, denoted as $\omega$. To ensure the lateral acceleration constraint $a_y \approx v^2 |\omega| < a_{l,\max}$ is satisfied, the vehicle speed must be dynamically limited.\\
To address these requirements, an effective reference speed $\bar{v}$ is introduced. This variable represents the most restrictive constraint among the desired speed, the regulatory speed limit, and the dynamic comfort limit derived from the instantaneous angular rate. The effective reference speed is mathematically defined as:
\begin{equation} \label{eq:v_bar_unified}
    \bar{v} = \min\left\{ v_{\text{des}}, v_{\max}, \sqrt{\frac{a_{l,\max}}{\left| \omega \right|}} \right\}\, \text{.}
\end{equation}
Consequently, a single quadratic cost $\mathbf{W}_v$ is utilized to penalize deviations from this calculated effective reference. This formulation ensures that the vehicle tracks the desired speed when it is possible, but seamlessly yields to comfort and safety constraints during high-curvature turns:
\begin{equation}\label{eq:speed_track_final}
    \mathbf{W}_v(v, \bar{v})=(v-\bar{v})^2\, \text{.}
\end{equation}
\vspace{-0.1cm}\subsection{Static APFs}\vspace{-0.3cm}
To ensure that the vehicle remains strictly within the road boundaries, repulsive APFs characterized by a Gaussian profile are established along the left ($l$) and right ($r$) road edges. Given a generic road boundary $i$ and the Euclidean distance of the vehicle from that boundary $h_i$, the potential field $\mathbf{W}_b^i(h_i)$ is defined as:
\begin{equation}\label{eq:rbnd}
    \textbf{W}_b^i(h_i)= e^{-(\eta h_i)^4}  \text{,}  \ \ i = l,r\, \text{.}
\end{equation}
The coefficient $\eta > 1$ is introduced to tune the smoothness of the repulsive potential's transition. The full road boundary potential $\mathbf{W}_b(h)$ is then obtained by the superposition of the left and right boundary APFs:
\begin{equation}\label{eq:rbnds}
    \textbf{W}_b(h) = \textbf{W}_b^l(h_l) + \textbf{W}_b^r(h_r)\, \text{.}
\end{equation}
This formulation guarantees that the vehicle remains within the road boundaries. However, it does not by itself ensure compliance with standard road regulations, which require keeping the rightmost free lane. To address this limitation, an additional sigmoid APF $\textbf{W}_l(h_c)$ is introduced. This term generates a virtual attractive force that guides the vehicle toward the preferred lane position. The APF is defined as a function of $h_c$, the minimum distance between the ego vehicle and the left boundary of the rightmost free lane. In a two-lane road scenario, $h_c$ reduces to the ego vehicle’s distance from the road centerline.
\begin{equation}\label{eq:APF_lane}
    \textbf{W}_l(h_c) = \frac{1}{(1+e^{h_c})}\, \text{.}
\end{equation}
\begin{remark}
The direct computation of Euclidean distances ($h_l$, $h_r$, $h_c$) to the road center line or boundaries could be computationally expensive in a Cartesian frame, since it requires iterative geometric procedures at runtime. A significant advantage of defining the problem within the Frenet reference frame (cf. Section~\ref{sec:fre}) is that these quantities are directly and analytically defined as simple functions of the lateral coordinate, $d$. Specifically, $h_l$, $h_r$, and $h_c$ are obtained through trivial algebraic operations on $d$ and the known lane/road geometry parameters.
\end{remark}\vspace{-.2cm}
Another pivotal aspect of safe driving is ensuring that the ego vehicle maintains the center of the occupied lane. As detailed in related works \citep{CaCeBo24}, it is not necessary to overly complicate the problem by introducing a new APF. A proper parameter computation, achieved by imposing suitable analytical conditions on the existing potential field, is used to obtain implicit lane centering. 
\vspace{-0.1cm}\subsection{Time-Varying APF}\vspace{-0.3cm}
A critical requirement for the LTP of AVs is the capability to account for the presence of dynamic obstacles characterized by uncertain future state evolution.
This uncertainty arises from distinct operational contexts. In the absence of V2X communication, the ego vehicle relies exclusively on on-board sensors, necessitating a conservative prediction strategy that covers the full range of reachable states for surrounding actors. Conversely, when V2X communication is available, the LTP benefits from richer intent information; however, it must still account for data uncertainty caused by communication latency, or inherent model-prediction errors in the received data.
\begin{assum} 
For the sake of clarity and consistent with standard road regulations, this work explicitly addresses uncertainties regarding actors' longitudinal position $s$ within the TVAPF definition, without loss of generality. The mathematical framework presented herein can be straightforwardly extended to lateral motion uncertainties.
\end{assum}
The prediction uncertainty associated with surrounding actors is formalized through the uncertain prediction vector $\Delta_i$ \eqref{eq:delta_i}. For a generic obstacle $i$ at a future time step $k+j$, predicted based on information available at the current time $k$, this vector is defined as:
\begin{equation}\label{eq:delta_i}
\Delta_i(k+j|k) = \begin{bmatrix}
\delta s_i(k+j|k) & s_{o,i}(k+j|k)
\end{bmatrix}^\top\, \text{.}
\end{equation}
This vector encapsulates two fundamental parameters: the estimated longitudinal uncertainty spread $\delta s$, and the predicted central longitudinal position of the obstacle $s_o$.\\
Given this uncertainty characterization, the TVAPF is defined to penalize collision risks dynamically over the prediction horizon:
\begin{align}\label{eq:apf_obs}
&\mathbf{W}_o(k+j|k) = e^{-\left(\left(\frac{s-s_o(k+j|k)}{\gamma_s(k+j|k)}\right)^c+\left(\frac{d-d_o(k)}{\gamma_d}\right)^c\right)} \ \text{,} \quad c\in \mathbb{N}_{\text{even}}^{+}\, \text{.} \nonumber\\
&\gamma_s(k+j|k)=\frac{\delta s(k+j|k)+\sigma_s}{\alpha_s}, \quad \gamma_d=\frac{l_W+\sigma_d}{\alpha_d} \, \text{.}
\end{align}
In this formulation, $\mathbf{W}_o$ represents the potential field magnitude at the predicted instant $k+j$. The parameters $\sigma_s$ and $\sigma_d$ denote the tunable safety margins that the ego vehicle is required to maintain longitudinally and laterally, respectively. The exponent $c$ governs the sharpness of the potential function's gradients, while $l_W$ defines the lateral occupancy of the obstacle field and is selected equal to the lane width. Finally, $\alpha_s$ and $\alpha_d$ are scaling factors which, following the methodology applied to lane potential fields, are computed to ensure that specific boundary conditions are satisfied at the limits of the safety zone.\\
In the following, the specific definition of TVAPF parameters is detailed in a scenario characterized by the absence of V2X communication. It is important to emphasize that this condition, while representing the most challenging operational scenario for the LTP, constitutes the necessary fallback procedure in the event of V2X subsystem faults.\\
Given the initial obstacle state at time $k$, defined by longitudinal position $s_o(k)$, lateral position $d_o(k)$, and speed $v_o(k)$, the predicted TVAPF parameters are computed for each step $j$ along the prediction horizon. The state evolution is propagated using the following discrete-time prediction model:
\begin{equation}\label{eq:actPred}
\begin{cases}
s_o(k+1) = s_o(k) + T_{s,L} v_o(k)\\
v_o(k+1) = v_o(k) + T_{s,L} a_o(k)
\end{cases}
\end{equation}
where $T_{s,L}$ denotes the sampling time of the LTP and $a_o$ the obstacle acceleration.\\
Due to the physical limitations of vehicle actors, speed and acceleration cannot vary arbitrarily but are constrained within the bounds $[v_{o,\min}, v_{o,\max}]$ and $[a_{o,\min}, a_{o,\max}]$, respectively. Consequently, the maximum and minimum reachable longitudinal positions, denoted as $s_{o,\max}(k+j|k)$ and $s_{o,\min}(k+j|k)$, are calculated for each prediction step by propagating these bounding dynamics and applying saturation constraints throughout the horizon.\\
To populate the uncertain prediction vector introduced in \eqref{eq:delta_i}, the longitudinal uncertainty spread $\delta s$ is defined as the spread of the reachable set at prediction step $k+j$:
\begin{equation}\label{eq:deltas}
\delta s (k+j|k) = \left|s_{o,\max}(k+j|k)-s_{o,\min}(k+j|k)\right|
\end{equation}
Similarly, the center of the uncertainty area is defined as the geometric center of the reachable set:
\begin{equation}\label{eq:so}
s_o(k+j|k)=\frac{s_{o,\max}(k+j|k)+s_{o,\min}(k+j|k)}{2}
\end{equation}
\begin{remark}\label{rm:un_compact}
The uncertain area defined by equations \eqref{eq:deltas} and \eqref{eq:so}, and the consequent TVAPF formulated in \eqref{eq:apf_obs}, constitute a \emph{compact bounded set} within the Frenet frame. This property is mathematically guaranteed by the inherent definition of the uncertainties, which are derived from physical constraints. Furthermore, due to the bounded dynamics governing vehicle motion described in \eqref{eq:actPred}, the time evolution and prediction of the potential field $\mathbf{W}_o$ are guaranteed to remain bounded over the entire prediction horizon.
\end{remark}
\vspace{-.2cm}
\section{Optimal Control Problem for Trajectory Planning}\vspace{-0.3cm}
In this section, we first define the prediction model utilized within the FHOCP.  Subsequently, the FHOCP formulation designed to unify DM and TP is detailed, and its key features are discussed.
\vspace{-0.1cm}\subsection{Prediction model}\vspace{-0.3cm}
Since the aim of the FHOCP is the generation of a feasible vehicle trajectory, the material point model reported in \eqref{simple_model_p} is utilized to account for the time behavior of the ego-vehicle:
\begin{equation} \label{simple_model_p}
\begin{cases}
    \dot{s}(t) = \nu(t)\cos(\psi(t)) \\
    \dot{d}(t) = \nu(t)\sin(\psi(t)) \\
    \dot{\psi}(t) = \omega(t) \\
    \dot{\nu}(t) = \alpha(t)
\end{cases}
\end{equation}
where $s$ and $d$ represent the vehicle's longitudinal and lateral positions in the Frenet frame, respectively.  The variable $\psi$ denotes the heading angle, $\nu$ represents the longitudinal speed, while $\alpha$ and $\omega$ correspond to the longitudinal acceleration and heading rate control inputs, respectively. To derive the discrete-time model required for the FHOCP predictions, the state vector is defined as $\xi = [s, d, \psi, \nu]^\top$ and the control input vector as $\lambda = [\alpha, \omega]^\top$. The prediction model \eqref{eq:simple_model_d_1} is then obtained by discretizing the continuous dynamics \eqref{simple_model_p} using the zero-order hold method with a sampling time of $T_{s,L}$.
\begin{equation}\label{eq:simple_model_d_1}
\xi(k+1) = f(\xi(k), \lambda(k))
\end{equation}
\vspace{-0.5cm}\subsection{Optimal Control Problem}\label{sec:ocp}\vspace{-0.3cm}
The LTP problem is cast as an FHOCP defined as a function of the uncertainties collected in the parameter matrix $\Delta$, the system control input sequence $\Lambda$, and the consequent predicted plant state $\Xi$.\\
The matrices collecting the necessary information over the prediction horizon $N_L$ are defined as follows:
\begin{enumerate}[leftmargin=*]
    \item Uncertainties Matrix $\Delta(k)$: collects the predicted uncertainties for all $n_o$ obstacles. It is defined as the matrix formed by the vectors $\Delta_i(k+j|k)$, where the row index $i \in \mathbb{N}_{n_o}$ corresponds to the obstacle and the column index $j \in \mathbb{N}_{N_L}$ corresponds to the prediction step.
    \item Control Input Sequence $\Lambda(k)$: the vector containing the sequence of optimized control inputs, defined as $\Lambda(k)= [\lambda(k|k)^\top, \dots, \lambda(k+N_L-1|k)^\top]^\top$. 
    \item Predicted State Sequence $\Xi(k)$: the vector containing the predicted state evolution, defined as $\Xi(k)= [\xi(k+1|k)^\top, \dots, \xi(k+N_L|k)^\top]^\top$.
\end{enumerate}
The cost function~\eqref{eq:cost} used in the FHOCP is composed of different objectives that penalize undesirable states or control actions, each multiplied by a proper weighting factor. The objectives previously described (e.g., function of speed, distances) can now be explicitly defined as functions of the optimization variables and parameters ($\Lambda, \Xi, \Delta$).
The FHOCP objective terms included in \eqref{eq:cost} are:
\begin{itemize}[leftmargin=*]
    \item $\mathbf{W}_v(\Xi)$: with cost weight $K_v$, this term penalizes deviations from the desired speed profile.
    \item $\mathbf{W}_b(\Xi)$: with cost weight $K_b$, this term enforces the vehicle to stay within lane boundaries.
    \item $\mathbf{W}_l(\Xi)$: with cost weight $K_l$, this term guides the vehicle toward the rightmost free lane.
    \item $\mathbf{W}_c(\Xi,\Lambda)$: with cost weight $K_c$, this term ensures proper passenger comfort by penalizing high lateral acceleration.
\end{itemize}
Finally, the term $\mathbf{O}(\Delta, \Xi)$ encompasses the contribution of the TVAPFs, which are crucial to account for the prediction uncertainties of surrounding actors. As defined in~\eqref{eq:apf_obs}, this term encapsulates the influence of all uncertainties related to the surrounding vehicles, integrating available information from various sources (e.g., V2X, sensors).
Following Remark~\ref{rm:un_compact}, the total uncertainty area defined by $\mathbf{O}$ is a finite sum of compact bounded sets, ensuring that $\mathbf{O}$ itself is compact and bounded.\\
For brevity, the detailed dependence on $\Lambda$, $\Xi$, and $\Delta$ will be implied in the formulation of the total cost function $\mathcal{J}$. The complete cost function $\mathcal{J}$ to be minimized is defined over the prediction horizon $N_L$ as:
\begin{align}
\mathcal{J}(\Lambda,\Xi,\Delta) = \sum_{j=0}^{N_L} &\left(K_v\textbf{W}_v + K_b\textbf{W}_b +  \right.\label{eq:cost}\\
\nonumber& \left. K_l\textbf{W}_l + K_c\textbf{W}_c + K_o\textbf{O} \right) \, \text{,}\\
   \textbf{O}(j) =&\ \sum_{i=1}^{n_o} \textbf{W}_{o,i}(\Delta_i(k+j|k)) \, \text{.}\label{eq:tvapf1}
\end{align}
The optimal vehicle trajectory is defined as the computed optimal state sequence $\Xi^*$ that minimizes the cost function~\eqref{eq:cost}. Although the proposed optimization setup may appear similar to standard MPC formulations, the primary goal in the proposed approach is different. The objective is not to evaluate the optimal control action $\lambda(k|k)$ at the current time instant, but to determine the entire state prediction $\Xi^*$ over the horizon $N_L$, which defines the optimal trajectory to be tracked by the MC. The trajectory $\Xi^*$ is finally the direct output of LTP.
\begin{equation}
\label{eq:nlp_complete}
    \Xi^* = \argmin_{\Lambda(k), \Xi(k)}  \mathcal{J}(\Lambda,\Xi,\Delta)
\end{equation}
\begin{equation*} 
\begin{aligned}
    & \text{subject} \text{ to}: \\
    & \xi(k|k) = \xi(k)\\
    & \xi(k+j+1|k)= f(\xi(k+j|k),\lambda(k))\, \text{,} && \forall j \in \mathbb{N}_{N_L-1}\, \text{,}\\
    & \xi(k+j|k) \in \mathcal{A}_\xi\, \text{,} && \forall j \in \mathbb{N}_{N_L}\, \text{,}\\
    & \lambda(k+j|k) \in \mathcal{A}_\lambda \, \text{,}&& \forall j \in \mathbb{N}_{N_L-1}\, \text{,}\\
    & \Delta \lambda(k+j|k) \in \mathcal{A}_{\Delta_\lambda} \, \text{,} && \forall j \in \mathbb{N}_{N_L-2}\, \text{,}\\
    & \mathbf{O}(\Delta,j) \leq \varepsilon_o\, \text{,} && \forall j \in \mathbb{N}_{N_L}\, \text{,}\\
    & \xi(k+N_L|k)\in \mathcal{A}_{\xi,f}\, \text{.}
\end{aligned}
\end{equation*}
To ensure trajectory feasibility, different constraints are introduced on the system dynamics and operational limits.  \\
Firstly, the plant state evolution is constrained by the actual vehicle state and must strictly adhere to the discrete prediction model~\eqref{eq:simple_model_d_1} over the entire horizon. To ensure compliance with both actuator feasibility limits and passenger comfort requirements, admissible sets are formally defined for the system variables. Specifically, we introduce $\mathcal{A}_\xi$ for the state vector, $\mathcal{A}_\lambda$ for the control inputs, and $\mathcal{A}_{\Delta_\lambda}$ for their respective rates of variation.\\
Finally, to guarantee safety, the vehicle's predicted position is constrained by the obstacle uncertainty TVAPF. This constraint ensures that the vehicle never crosses a region that could potentially be occupied by other actors over the prediction horizon. Consequently, the obtained optimal trajectory $\Xi^*$ is safe despite the uncertain behavior of the surrounding actors.\\
Recursive feasibility guarantees that if a solution exists at time $k$, a feasible solution continues to exist at any subsequent time $k+n$. This property is fundamental for online implementation and relies on specific consistency requirements between the planning and control layers.
\begin{assum}[Hierarchical Consistency]
\label{ass:consistency}
The interaction between the LTP and the MC is assumed to satisfy two conditions. Firstly, the MC feasible input set $\mathcal{A}_\mathrm{u}$ is a strict subset of the corresponding LTP constraints. Secondly, the MC tracks the LTP reference trajectory with a bounded error contained within the set $\mathcal{A}_\mathrm{e}$.
\end{assum} \vspace{-.2cm}
Under Assumption~\ref{ass:consistency}, the recursive feasibility of the FHOCP could be ensured by the design of the terminal set. Rather than imposing an overly restrictive complete stop at the end of the prediction horizon, which would severely limit performance, a \emph{Safe Stop} paradigm is adopted. The terminal state is constrained to lie within a specific region $\mathcal{A}_{\xi,\mathrm{f}}$ from which a collision-free braking maneuver is guaranteed to exist.\\
This region is determined by calculating a dynamic longitudinal stopping distance $\mathcal{D}(\nu_{\text{ter}})$, which accounts for the system reaction time $\tau$, the distance covered during jerk build-up ($j_{\max}$), and the constant deceleration phase ($\alpha_{\min}$):
\begin{equation}\label{eq:a_braking_dist}
    \mathcal{D}(\nu_{\text{ter}}) = \nu_{\text{ter}} \left( \tau + \frac{|\alpha_{\min}|}{j_{\max}} + \frac{\nu_{\text{ter}}}{2|\alpha_{\min}|} \right) \, \text{.}
\end{equation}
where $v_{\text{ter}}$ is the maximum allowable terminal speed. By defining the safe longitudinal limit $s_t = s_{\text{o}}(k+N_L|k) - \delta s(k+N_L|k) - \mathcal{D}(\nu_{\text{ter}})$, the terminal invariant set is defined as:
\begin{equation}\label{eq:terminal_set}
    \mathcal{A}_{\xi,\mathrm{f}} =  \left\{ \xi : \, 
    \begin{aligned}
    & 0 \leq s(k+N_L|k) \leq s_t, \\
    & |d(k+N_L|k) - d_{\text{center}}| \leq \varepsilon_d, \\
    & |\psi(k+N_L|k)| \leq \varepsilon_{\psi}, \\
    & 0 \leq \nu(k+N_L|k) \leq \nu_{\text{ter}}
    \end{aligned}
    \right\} \, \text{.}
\end{equation}
where $\varepsilon_d$ and $\varepsilon_\psi$ are small relaxation tolerances to guarantee the numerical feasibility and $d_{\text{center}}$ represents the lateral coordinate of the lane center.
The recursive feasibility is guaranteed by the construction of this set. Since any state $\xi \in \mathcal{A}_{\xi,\mathrm{f}}$ admits a valid feedback control law (specifically, the maximum braking maneuver defined by $\mathcal{D}$) that keeps the system trajectories collision-free and drives the vehicle to a \emph{Safe Stop} (an invariant state), a feasible solution for the FHOCP always exists by extending the current optimal trajectory with this safety maneuver.
\begin{exmp}
Fig.~\ref{fig:TVAPF} illustrates the effectiveness of the TVAPF in unifying DM and trajectory planning within a highway scenario involving a leading vehicle (A) and an approaching vehicle on the left lane (B). The snapshots depict the TVAPF evolution over the prediction horizon for a single LTP instance. In the top scenario, the LTP successfully computes an overtaking maneuver, as evidenced by the ego vehicle avoiding the uncertain regions of the TVAPF. Conversely, in the bottom scenario, the proximity of vehicle B renders the overtake unsafe; consequently, the LTP concurrently \emph{decides} to keep the acutal lane and \emph{plans} a coherent lane-keeping trajectory to maintain a safe distance. Although this example highlights a single time step, the FHOCP is solved iteratively at runtime to continuously adapt to environmental state updates and uncertainties.
\begin{figure}[]
    \centering
    \begin{minipage}[b]{0.45\textwidth}
        \centering
        \includegraphics[width=\textwidth]{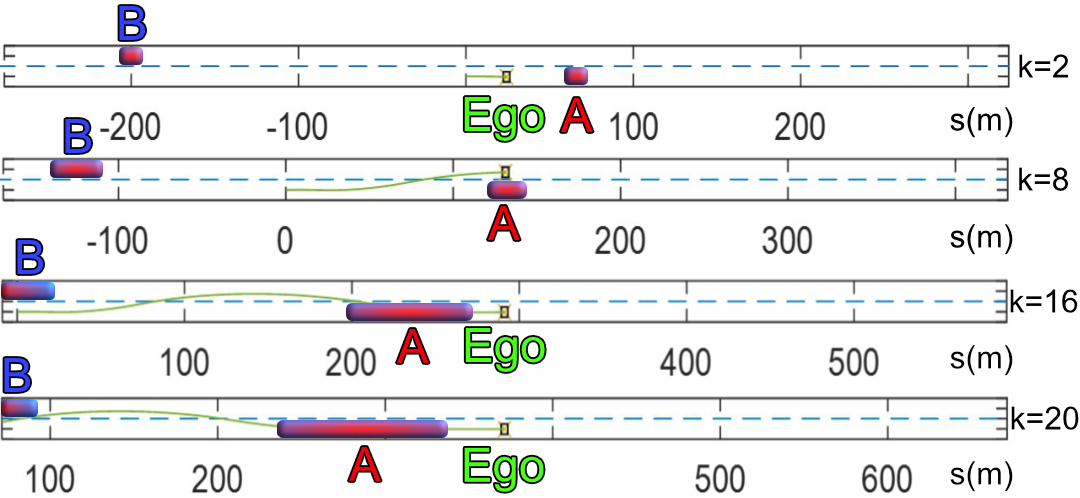}
    \end{minipage}
    \begin{minipage}[b]{0.45\textwidth}
        \centering
        \includegraphics[width=\textwidth]{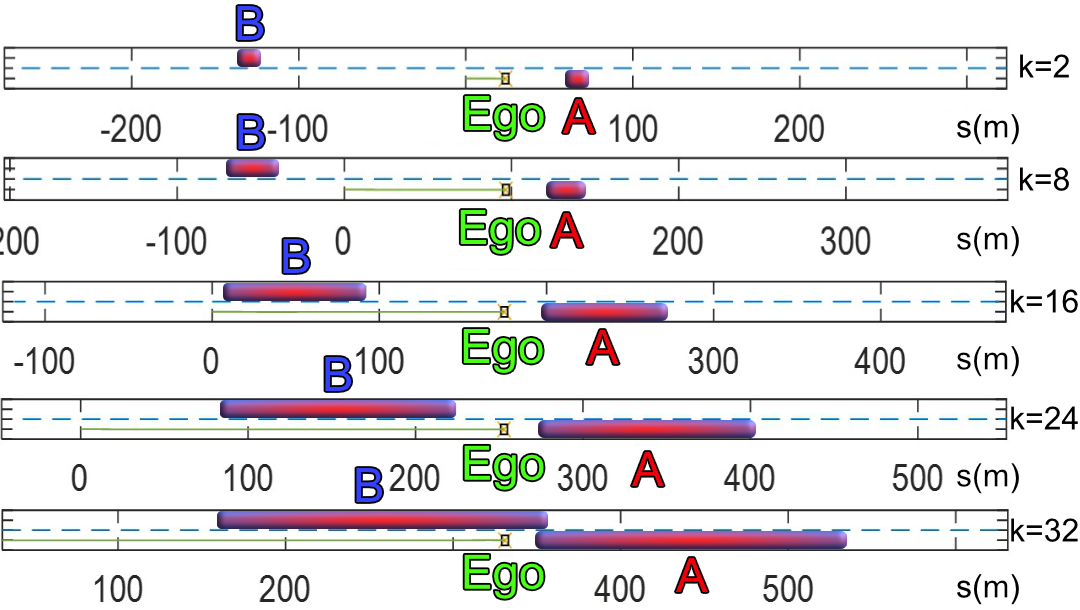}
    \end{minipage}
    \caption{TVAPF prediction in the LTP FHOCP.\\ Top: Feasible overtaking maneuver. Bottom: The LTP detects collision risk, deciding to maintain the current lane behind vehicle A.}
    \label{fig:TVAPF}
\end{figure}
\end{exmp}
\vspace{-0.3cm}
\section{Motion controller}\vspace{-0.3cm}
An NMPC scheme is proposed to track the LTP reference trajectory. While serving as an illustrative implementation, the framework is compatible with alternative MCs that satisfies Assumption~\ref{ass:consistency}. The NMPC employs a single-track kinematic model, which is proven reliable for non-high-speed operations \cite{stkLS}.  The system evolution is governed by the following continuous-time differential equations:
\begin{equation}
    \begin{cases}
        \dot{x}(t) = v(t)\cos(\psi(t))\\
        \dot{y}(t) = v(t)\sin(\psi(t))\\
        \dot{\theta}(t) = \frac{v(t)}{L} \tan(\delta(t))\\
        \dot{v}(t) = a(t)\\
        \dot{\delta}(t) = w_{\delta}(t)
    \end{cases}\, \text{.}
\end{equation}
where $x, y$ denote the rear-axle position, $\theta$ the heading, $v$ the velocity, $\delta$ the steering angle and $L$ the vehicle wheelbase. The control inputs are the longitudinal acceleration $a$ and the steering rate $w_\delta$. Accordingly, the state and input vectors are defined as $\chi = [x, y, \psi, v, \delta]^\top$ and $u = [a, w_\delta]^\top$. Discretization with sampling time $T_{s,\text{MPC}}$ yields the discrete-time model:
\begin{equation}\label{stk_d}
    \chi(k+1) = g(\chi(k), u(k))\, \text{.}
\end{equation}
The tracking error at step $k$ is defined with respect to the reference trajectory $\chi_{\text{ref}}$:
\begin{equation}
    e(k) = \chi(k) - \chi_{\text{ref}}(k)\, \text{.}
\end{equation}
The tracking problem is cast as a constrained FHOCP at each discrete time instance $k$. The control objective is to minimize the quadratic cost function $\mathcal{J}_{\text{MPC}}$ \eqref{eq:cost_mpc}, which penalizes the deviation from the reference trajectory and the control effort over the prediction horizon $N_P$, utilizing weight matrices $Q$ and $R$, respectively. The optimization yields the optimal control sequence $U^*(k)$ by minimizing this tracking cost augmented with a penalty $\rho$ on the slack variable $\sigma$:
\begin{align}
    \mathcal{J}_{\text{MPC}}(k) =& \sum_{j=0}^{N_P-1} \left( \lVert e(k+j|k)\rVert ^2_{Q} + \lVert u(k+j|k)\rVert^2_{R} \right) \text{.}\label{eq:cost_mpc} \\
    U^*(k), \sigma^* = & \ \argmin_{U(k), \sigma} \left( \mathcal{J}_{\text{MPC}}(k) + \rho \sigma^2 \right)\, \text{.} \label{eq:argmin_mpc}
\end{align}
\begin{equation*} 
    \begin{aligned}
        & \text{subject to:}\\
        & \chi(k|k) = \chi(k) \, \text{,}\\
        & \chi(k+j+1|k)= g(\chi(k+j|k), u(k+j|k)), && \forall j \in \mathbb{N}_{N_P-1}\, \text{,}\\
        & e(k+j|k) \in \mathcal{A}_e, && \forall j \in \mathbb{N}_{N_P}\, \text{,}\\
        & u(k+j|k) \in \mathcal{A}_u, && \forall j \in \mathbb{N}_{N_P-1}\, \text{,}\\
        & \Delta u(k+j|k) \leq \mathcal{A}_{\Delta u}, && \forall j \in \mathbb{N}_{N_P-2}\, \text{,}\\
        & \lVert e(k+N_P|k) \rVert_{I}^2 \leq \sigma \, \text{,} \qquad   \sigma \geq 0 \, \text{.}
    \end{aligned}
\end{equation*}
The NMPC optimization is constrained by the discrete-time dynamics \eqref{stk_d}. Additionally, the admissible sets for the states and inputs are defined to strictly satisfy Assumption~\ref{ass:consistency}, and the input rate is bounded within $\mathcal{A}_{\Delta u}$. Furthermore, to guarantee the nominal asymptotic stability of the closed-loop system, a terminal equality constraint on the tracking error is enforced (c.f. \cite{book_mpc}). In this formulation, the introduction of the slack variable $\sigma$ effectively converts this into a soft constraint, preventing numerical infeasibility while ensuring the error is driven to zero.
\section{Simulation Results}\vspace{-0.3cm}
In this section, we introduce a simulation test to show the effectiveness of the proposed approach. 
\vspace{-0.1cm}\subsection{Test Environment}\vspace{-0.3cm}
The simulation test is developed via MatLab and Simulink tools. The \textit{Roadrunner} environment (\cite{mathworks_roadrunner}) is used to define the driving scenario, and the \textit{Vehicle Dynamics Blockset} is employed to simulate vehicle dynamics. In this setup, only the actual state of the surrounding vehicles in the ego vehicle sensor range is known. The relevant optimization problems \eqref{eq:nlp_complete} \eqref{eq:argmin_mpc} involved in the LTP and the NMPC are solved with the \textit{Ipopt} solver (\cite{wachter2006ipopt}) using CasADi (\cite{andersson2019casadi}) as a development framework. The parameters used for the OCPs are reported in Table \ref{tab:ltp_nmpc_fhocp}. In all simulations performed, the maximum LTP optimization time resulted in $0.23$ s while the mean in $0.031$ s on an Intel i5-1340P CPU.
\begin{table}[]
    \centering
    \caption{LTP and NMPC FHOCP parameters}
    \label{tab:ltp_nmpc_fhocp}
    \begin{tabular}{l c c}
        \hline
        Parameter & LTP & NMPC \\
        \hline
        Sample time $T_s$ & 0.5 \(\mathrm{s}\)  & 0.2 \(\mathrm{s}\) \\
        Prediction horizon     & 35 \(\mathrm{s}\) & 2 \(\mathrm{s}\) \\
        Instance period    & 5 \(\mathrm{s}\) & 0.2 \(\mathrm{s}\)\\
        \hline
    \end{tabular}
\end{table}
\begin{figure}
    \centering
    \includegraphics[width=0.85\linewidth,trim={0 1.5cm 0 2.5cm},clip]{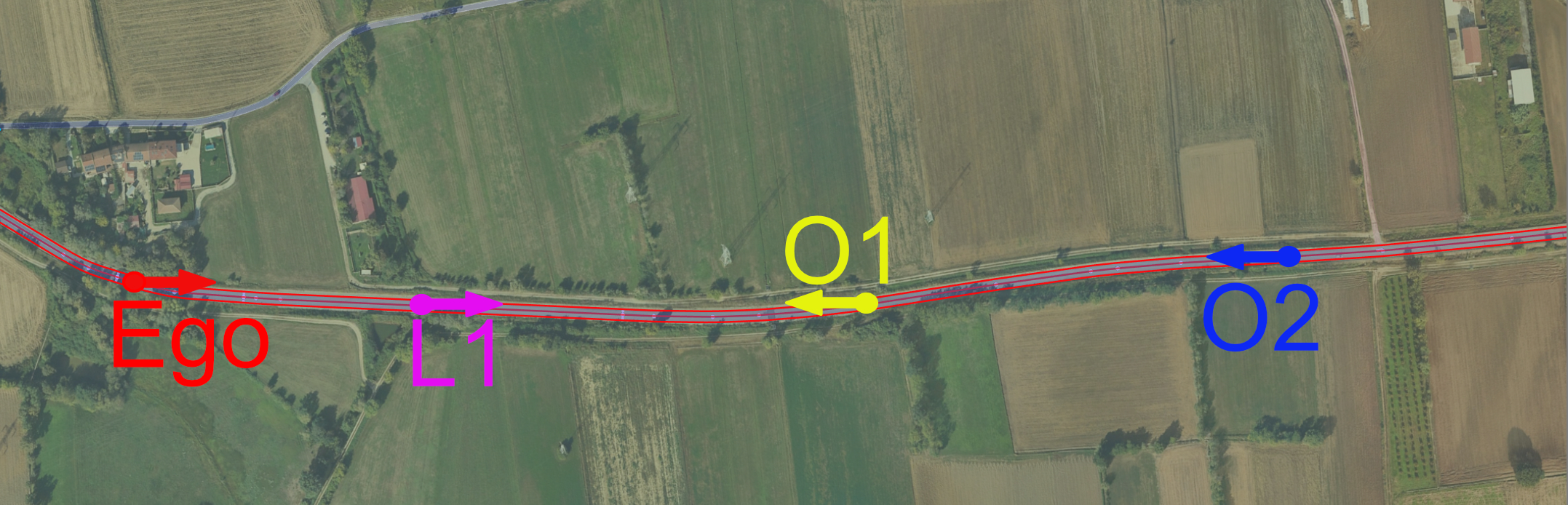}
    \caption{Overview of the scenario in Roadrunner, detailing the position of the ego vehicle and the actors (L1,O1,O2) before the overtake maneuver.}
    \label{fig:scenario-over}
\end{figure}
\vspace{-0.1cm}\subsection{Test Description}
\begin{figure*}[]
  \centering
  \includegraphics[width=0.82\textwidth]{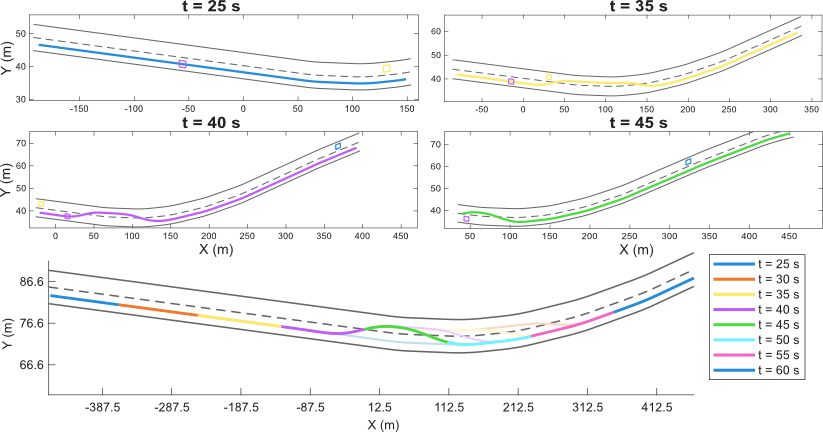}
  \caption{LTP trajectories: 
  Top: Predicted trajectory at a specific LTP instance, showing the current position of the actors. Bottom: Trajectory segments followed by the ego vehicle, including semi-transparent extensions that represent the complete predicted path at each LTP instance.}
  \label{fig:Overtake}
\end{figure*}

\begin{figure} 
    \centering
    \includegraphics[width=0.9\linewidth]{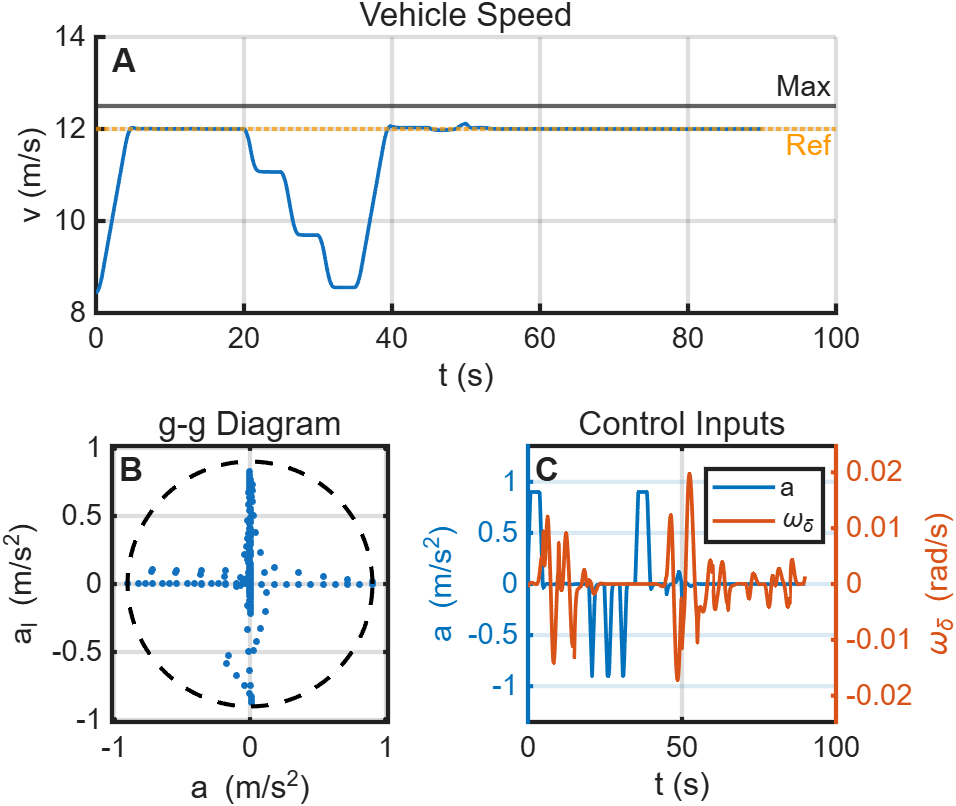}
    \caption{Simulation results: A - Vehicle speed. B - g-g diagram. C - Control Inputs.}
    \label{fig:Scenario3}
\end{figure}

\vspace{-0.3cm}
For the simulation test\footnote{An animation video of the simulation is available online \url{https://youtube.com/playlist?list=PLaJBGNRMGh8KrlUwjy9TqowIFCURIfwDC&si=oSGsbinLMCf03ize}.}, we consider the driving scene represented in Fig. \ref{fig:scenario-over} consisting of a suburban two-way road with lane width $l_W=4$ m and maximum speed limitation of $12.5$ m/s. In this scene, besides the ego vehicle, three actors are present: L1 traveling ahead of the ego in the same direction and, in the other lane, the actors O1, O2 moving in the opposite direction. All vehicles are considered subject to the constraints in Table~\ref{tab:limits}.\\
\begin{table}[]
    \centering
    \caption{Vehicles kinematic limits}
    \label{tab:limits}
    \begin{tabular}{l c c c}
        \hline
        Parameter & Unit & Minimum & Maximum \\
        \hline
        Speed \(v\)  & \(\mathrm{m/s}\)  & 0 & 12.5   \\
        Longitudinal acceleration \(a\) & \(\mathrm{m/s^2}\)  & -0.9 & 0.9\ \\
        Jerk \(\dot{a}\) & \(\mathrm{m/s^3}\)  & -0.9 & 0.9\ \\
        Yaw rate \(\dot{\theta}\)  & \(\mathrm{deg/s}\)& -4.44 & 4.44\\
        Steering angle \(\delta\) & \(\mathrm{deg}\) & -24.5 & 24.5\\
        \hline
    \end{tabular}
\end{table}
The scenario starts with the ego vehicle entering a 1.5 km road segment at an initial speed of 8.33m/s, with a reference velocity of 12 m/s. The lead vehicle L1 is located 300 m ahead, traveling at a lower speed of 3 m/s (see Fig.~\ref{fig:scenario-over}) and, in this condition, an overtaking maneuver has to be performed by the ego vehicle. 
The scenario is characterized by the dynamic behavior of the surrounding actors. In fact, as the ego vehicle closes the gap to L1, the oncoming vehicle O1 approaches (Fig.~\ref{fig:scenario-over}) and accelerates with $a_{\max}$ from 6 m/s to 10 m/s. Simultaneously, the lead vehicle L1 accelerates up to 6 m/s. Furthermore, the second oncoming vehicle O2, located behind O1, approaches at a constant speed of 8 m/s.
Under these dynamic conditions, the LTP must adapt the vehicle speed to ensure collision avoidance. Concurrently, it must determine the optimal timing and trajectory for a physically feasible and comfortable overtaking.\\
In particular, the simulation starts with the ego vehicle accelerating to reach the reference speed of $12.5$ m/s (Fig. \ref{fig:Scenario3}~A). At 15 s, the LTP receives information about the position and velocity of L1 and O2, and TVAPF are computed to predict their worst case behavior. Thus, in the next three LTP instances (at 20 s, 25 s, and 30 s), the LTP decides to delay the overtake maneuver as all overtake trajectories are rendered infeasible by the TVAPF. Consequently, the LTP progressively reduces the ego speed till 8.6 m/s to keep a safe distance from L1. Furthermore, Fig. \ref{fig:Scenario3} A shows that the deceleration speed profile directly results from the sequence of LTP instances, rather than from a static velocity target.
In the next LTP instance, at 30 s, even the worst case prediction encoded by the TVAPF allows the ego to perform the overtake maneuver, and an initial overtaking path is computed.\\
In the successive LTP instances at 35 s, 40 s, and 45 s, the feasible overtaking path is progressively refined (Fig. \ref{fig:Overtake}), accounting for the arrival of another oncoming vehicle in the left lane (O2) and by the usage of new information, that reduces the uncertainty in the predicted positions of O1 and L1. Finally, at 55 s, the ego has completed the maneuver and returned to the rightmost lane with no leading vehicle, thus it seamlessly resumes tracking the lane center at the reference speed (Fig. \ref{fig:Scenario3}).\\
In summary, despite the uncertain dynamics of the surrounding actors, the proposed architecture effectively manages the timing of the overtaking maneuver. Furthermore, it ensures the generation of a safe trajectory that is continuously reshaped as a function of the most accurate information available at each planning instance. During the entire simulation, passenger comfort is also preserved, as illustrated by the g–g diagram in Fig. \ref{fig:Scenario3}~B, where both longitudinal and lateral accelerations remain within the imposed limits.
\vspace{-0.1cm}
\subsection{Discussion}
\vspace{-0.3cm}
The proposed simulation highlights specific features of the discussed unified DM-TP approach. 
In particular, the sequence of the made decisions does not rely on a FSM with a predefined set of driving modes (such as distance keeping, overtaking, etc.) and a subsequent trajectory generation strategy (e.g., splines, clothoids, etc.) (c.f. \cite{L3CaRa}). Therefore, this formulation guarantees consistency between the strategic decision and the existence of a physically feasible path. This ensures that the generated trajectory is inherently adaptive to the evolving scenario. Additionally, the approach overcomes common FSM limitations, such as undefined states or conflicting transition conditions. Finally, it strictly maintains state and control continuity, avoiding the complexities associated with discrete mode switching.
\vspace{-0.05cm}
\section{Conclusion}\vspace{-0.3cm}
This paper presents a unified framework for the DM-TP of AVs, exploiting a novel TVAPF formulation. By explicitly modeling the future state evolution and bounded uncertainties of dynamic obstacles, the proposed method seamlessly integrates information from both on-board perception and V2X information.\\ 
The embedding of these dynamic fields within a tailored FHOCP enables the simultaneous evaluation of the most suitable driving maneuver and the consequent generation of a physically feasible trajectory. 
Simulation results in a dynamic multi-actor scenario characterized by real road topology demonstrated the effectiveness of the approach. The framework successfully generates strategically coherent behaviors while satisfying safety and comfort constraints. Future work will focus on extending the architecture to more complex scenarios, with particular emphasis on intersection handling.

\bibliography{ifacconf}

@ARTICLE{L3CaRa,
  author={Canale, Massimo and Razza, Valentino},
  journal={IEEE Access}, 
  title={Automated Driving Control in Highway Scenarios Through a Two-Level Hierarchical Architecture}, 
  year={2024},
  volume={12},
  number={},
  pages={86470-86486},
  doi={10.1109/ACCESS.2024.3416670}}

@article{SWctrl,
author = {Wang, Xue-Fang and Chen, Wen-Hua and Jiang, Jingjing and Yan, Yunda},
title = {High-level decision-making for autonomous overtaking: An MPC-based switching control approach},
journal = {IET Intelligent Transport Systems},
volume = {18},
number = {7},
pages = {1259-1271},
doi = {https://doi.org/10.1049/itr2.12507},
year = {2024}
}

@ARTICLE{ColAvoid,
  author={Yang, Hongjiu and He, Yongqi and Xu, Yang and Zhao, Hai},
  journal={IEEE Transactions on Intelligent Vehicles}, 
  title={Collision Avoidance for Autonomous Vehicles Based on MPC With Adaptive APF}, 
  year={2024},
  volume={9},
  number={1},
  pages={1559-1570},
  doi={10.1109/TIV.2023.3337417}
  }

@article{Bezier,
author = {Zheng, Ling and Zeng, Pengyun and Yang, Wei and Li, Yinong and Zhan, Zhenfei},
title = {Bézier curve-based trajectory planning for autonomous vehicles with collision avoidance},
journal = {IET Intelligent Transport Systems},
volume = {14},
number = {13},
pages = {1882-1891},
doi = {https://doi.org/10.1049/iet-its.2020.0355},
year = {2020}
}

@INPROCEEDINGS{AAPF,
  author={Lin, Pengfei and Tsukada, Manabu},
  booktitle={2022 13th Asian Control Conference (ASCC)}, 
  title={Adaptive Potential Field with Collision Avoidance for Connected Autonomous Vehicles}, 
  year={2022},
  volume={},
  number={},
  pages={2251-2256},
  doi={10.23919/ASCC56756.2022.9828160}
}

@ARTICLE{AAPF1,
  author={Lu, Bing and Li, Guofa and Yu, Huilong and Wang, Hong and Guo, Jinquan and Cao, Dongpu and He, Hongwen},
  journal={IEEE Access}, 
  title={Adaptive Potential Field-Based Path Planning for Complex Autonomous Driving Scenarios}, 
  year={2020},
  volume={8},
  number={},
  pages={225294-225305},
  doi={10.1109/ACCESS.2020.3044909}}

@INPROCEEDINGS{ChLiYa14,
  author={Cheng, Liping and Liu, Chuanxi and Yan, Bo},
  booktitle={2014 IEEE International Conference on Information and Automation (ICIA)}, 
  title={{Improved hierarchical A-star algorithm for optimal parking path planning of the large parking lot}}, 
  year={2014},
  volume={},
  number={},
  pages={695-698},
  doi={10.1109/ICInfA.2014.6932742}
}

@ARTICLE{4082128,
  author={Hart, Peter E. and Nilsson, Nils J. and Raphael, Bertram},
  journal={IEEE Transactions on Systems Science and Cybernetics}, 
  title={A Formal Basis for the Heuristic Determination of Minimum Cost Paths}, 
  year={1968},
  volume={4},
  number={2},
  pages={100-107},
  doi={10.1109/TSSC.1968.300136}}

@INPROCEEDINGS{CaCeBo24,
  author={Canale, Massimo and Cerrito, Francesco and Borodani, Pandeli},
  booktitle={2024 IEEE 63rd Conference on Decision and Control (CDC)}, 
  title={An Ego-Based Approach to Planning and Control for Automated Valet Parking Applications}, 
  year={2024},
  volume={},
  number={},
  pages={8193-8198},
  doi={10.1109/CDC56724.2024.10886152}}

@misc{mathworks_roadrunner,
  author       = {{MathWorks}},
  title        = {RoadRunner},
  subtitle     = {3D Scene Editor for Automated Driving Systems},
  year         = {2024}
}

@article{andersson2019casadi,
  author  = {Andersson, Joel A. E. and Gillis, Joris and Horn, Greg and Rawlings, James B. and Diehl, Moritz},
  title   = {{CasADi} -- {A} software framework for nonlinear optimization and optimal control},
  journal = {Mathematical Programming Computation},
  year    = {2019},
  volume  = {11},
  number  = {1},
  pages   = {1--36},
  doi     = {10.1007/s12532-018-0139-4}
}

@article{wachter2006ipopt,
  author  = {W{\"a}chter, Andreas and Biegler, Lorenz T.},
  title   = {On the implementation of an interior-point filter line-search algorithm for large-scale nonlinear programming},
  journal = {Mathematical Programming},
  year    = {2006},
  volume  = {106},
  number  = {1},
  pages   = {25--57},
  doi     = {10.1007/s10107-004-0559-y}
}

@article{APF0,
author = {Oussama Khatib},
title ={Real-Time Obstacle Avoidance for Manipulators and Mobile Robots},
journal = {The International Journal of Robotics Research},
volume = {5},
number = {1},
pages = {90-98},
year = {1986},
doi = {10.1177/027836498600500106},
}

@INPROCEEDINGS{APF_std0,
  author={Zhu, Qidan and Yan, Yongjie and Xing, Zhuoyi},
  booktitle={Sixth International Conference on Intelligent Systems Design and Applications}, 
  title={Robot Path Planning Based on Artificial Potential Field Approach with Simulated Annealing}, 
  year={2006},
  volume={2},
  number={},
  pages={622-627},
  doi={10.1109/ISDA.2006.253908}}

@Article{APF_std1,
author={Rostami, Seyyed Mohammad Hosseini
and Sangaiah, Arun Kumar
and Wang, Jin
and Liu, Xiaozhu},
title={Obstacle avoidance of mobile robots using modified artificial potential field algorithm},
journal={EURASIP Journal on Wireless Communications and Networking},
year={2019},
month={Mar},
day={18},
volume={2019},
number={1},
pages={70},
issn={1687-1499},
doi={10.1186/s13638-019-1396-2},
}

@INPROCEEDINGS{stkLS,
  author={Kong, Jason and Pfeiffer, Mark and Schildbach, Georg and Borrelli, Francesco},
  booktitle={2015 IEEE Intelligent Vehicles Symposium (IV)}, 
  title={Kinematic and dynamic vehicle models for autonomous driving control design}, 
  year={2015},
  volume={},
  number={},
  pages={1094-1099},
  doi={https://doi.org/10.1109/IVS.2015.7225830}
}

@ARTICLE{FSM1,
  author={Zheng, Xunjia and Li, Huilan and Zhang, Qiang and Liu, Yonggang and Chen, Xing and Liu, Hui and Luo, Tianhong and Gao, Jianjie and Xia, Lihong},
  journal={Journal of Intelligent and Connected Vehicles}, 
  title={Intelligent Decision-Making Method for Vehicles in Emergency Conditions Based on Artificial Potential Fields and Finite State Machines}, 
  year={2024},
  volume={7},
  number={1},
  pages={19-29},
  doi={10.26599/JICV.2023.9210025}
}

@article{survDM,
author = {S M Veres and L Molnar and N K Lincoln and C P Morice},
title ={Autonomous vehicle control systems — a review of decision making},
journal = {Proceedings of the Institution of Mechanical Engineers, Part I: Journal of Systems and Control Engineering},
volume = {225},
number = {2},
pages = {155-195},
year = {2011},
doi = {10.1177/2041304110394727}
}

@Inbook{book_mpc,
author="Gr{\"u}ne, Lars and Pannek, J{\"u}rgen",
title="Nonlinear Model Predictive Control",
bookTitle="Nonlinear Model Predictive Control: Theory and Algorithms",
year="2017",
publisher="Springer International Publishing",
address="Cham",
pages="45--69",
isbn="978-3-319-46024-6",
}

@ARTICLE{DMLTP,
  author={Liu, Wenru and Liu, Haichao and Zheng, Lei and Huang, Zhenmin and Ma, Jun},
  journal={IEEE Transactions on Vehicular Technology}, 
  title={Synergizing Decision Making and Trajectory Planning Using Two-Stage Optimization for Autonomous Vehicles}, 
  year={2025},
  volume={74},
  number={4},
  pages={5489-5503},
  doi={10.1109/TVT.2024.3509515}}

@article{DMLTP2,
author = {Tengfei Fu and Hongliang Zhou and Zhiyuan Liu},
title ={Integrated decision-making and path planning framework for autonomous driving in multi-lane obstacle avoidance},
journal = {Proceedings of the Institution of Mechanical Engineers, Part D: Journal of Automobile Engineering},
pages = {15},
year = {2025},
doi = {10.1177/09544070251327764},
}

@INPROCEEDINGS{survMC,
  author={Calzolari, Davide and Schürmann, Bastian and Althoff, Matthias},
  booktitle={2017 IEEE 20th International Conference on Intelligent Transportation Systems (ITSC)}, 
  title={Comparison of trajectory tracking controllers for autonomous vehicles}, 
  year={2017},
  volume={},
  number={},
  pages={1-8},
  doi={10.1109/ITSC.2017.8317800}}
        
\end{document}